\documentclass[aps,prl,floatfix,superscriptaddress,showpacs,twocolumn]{revtex4}

\usepackage{longtable}
\usepackage{bbm}
\usepackage{times}
\usepackage{nicefrac}
\usepackage{amsfonts}
\usepackage[ansinew]{inputenc}
\usepackage{epsfig}
\usepackage{graphicx}
\usepackage{color}
\usepackage{bm}
\usepackage{multirow}
\usepackage{mathrsfs}
\usepackage{revsymb}
\usepackage{amssymb}
\usepackage{amsmath}
\usepackage{mathtools}
\usepackage{slashed}
\usepackage{longtable}
\usepackage{dcolumn}
\usepackage[babel]{csquotes}

\newcolumntype{.}{D{x}{}{-1}}

\newcommand{\Za}{Z\alpha}

\begin{document}

\title{Theory of the two-loop self-energy correction to the $\bm g$ factor in non-perturbative Coulomb~fields}

\author{B. Sikora}
\email[]{bsikora@mpi-hd.mpg.de}
\address{Max~Planck~Institute for Nuclear Physics, Saupfercheckweg~1, D~69117 Heidelberg, Germany}
\author{V.~A. Yerokhin}
\address{Max~Planck~Institute for Nuclear Physics, Saupfercheckweg~1, D~69117 Heidelberg, Germany}
\address{Center for Advanced Studies, Peter the Great St.~Petersburg Polytechnic University, 195251 St.~Petersburg, Russia}
\author{N.~S. Oreshkina}
\address{Max~Planck~Institute for Nuclear Physics, Saupfercheckweg~1, D~69117 Heidelberg, Germany}
\author{H. Cakir}
\address{Max~Planck~Institute for Nuclear Physics, Saupfercheckweg~1, D~69117 Heidelberg, Germany}
\author{C.~H. Keitel}
\address{Max~Planck~Institute for Nuclear Physics, Saupfercheckweg~1, D~69117 Heidelberg, Germany}
\author{Z. Harman}
\email[]{harman@mpi-hd.mpg.de}
\address{Max~Planck~Institute for Nuclear Physics, Saupfercheckweg~1, D~69117 Heidelberg, Germany}

\begin{abstract}

Two-loop self-energy corrections to the bound-electron $g$ factor are investigated theoretically to all orders in the nuclear binding
strength parameter $\Za$. The separation of divergences is performed by dimensional regularization, and the contributing diagrams are
regrouped into specific categories to yield finite results. We evaluate numerically the loop-after-loop terms, and the remaining diagrams
by treating the Coulomb interaction in the electron propagators up to first order. The results show that such two-loop
terms are mandatory to take into account for projected near-future stringent tests of quantum electrodynamics and for the
determination of fundamental constants through the $g$ factor.

\end{abstract}


\maketitle

The $g$ factor of one-electron ions can be measured and calculated with an exceptional accuracy
\cite{Sturm11,Sturm13,Pachucki04,Pachucki2017,Yerokhin02,Shabaev02,Beier00,Karshenboim01,Lee05,Czarnecki2018}.
Its theoretical and experimental values in $^{28} \mathrm{Si}^{13+}$ were found to be in excellent agreement
\cite{Sturm11}. Since then, the experimental uncertainty decreased by an order of magnitude \cite{Sturm13}. Such measurements also allowed
an improved determination of the electron mass~\cite{Sturm14} (see also
\cite{Haffner00,Verdu04}). It is anticipated that bound-electron $g$ factor measurements will also enable in the foreseeable future an
independent determination of the fine-structure constant $\alpha$ \cite{Shabaev06,Yerokhin2016alpha}.

To push forward the boundaries of theory, quantum electrodynamic (QED) corrections at the one- and two-loop level need to be calculated with
increasing accuracy. One-loop corrections have been evaluated both as a power series in $Z \alpha$ (with $Z$ being the atomic number) and
non-perturbatively in this parameter (see e.g. \cite{Pachucki05,Yerokhin04,Yerokhin2017Oneloop}). Two-loop corrections were evaluated up to fourth
order in $\Za$ \cite{Pachucki05,Czarnecki16}. Contributions of order $\alpha^2 (\Za)^5$ were completed very recently \cite{Czarnecki2018}. At high
nuclear charges, where $\Za\approx 1$, an expansion in $\Za$ is not applicable. So far, the two-loop diagrams with
two electric vacuum polarization (VP) loops and those with one electric VP and one self-energy (SE) loop were evaluated non-perturbatively in
$Z \alpha$ \cite{Yerokhin2Loop2013}.

For a broad range of $Z$, the two-loop SE corrections, which are by far the hardest to calculate, constitute the largest
source of uncertainty. This holds true even at $Z=6$, after a recent high-precision evaluation of the one-loop SE corrections
\cite{Yerokhin2017Oneloop,Pachucki2017}. We thus see that higher-order terms in $\Za$ are also necessary at lower nuclear charges, if an ultimate
precision is required. Therefore, in the current Letter we present the theoretical framework for the non-perturbative evaluation of the two-loop
SE terms.

\begin{figure}[t]
\includegraphics[width=\columnwidth]{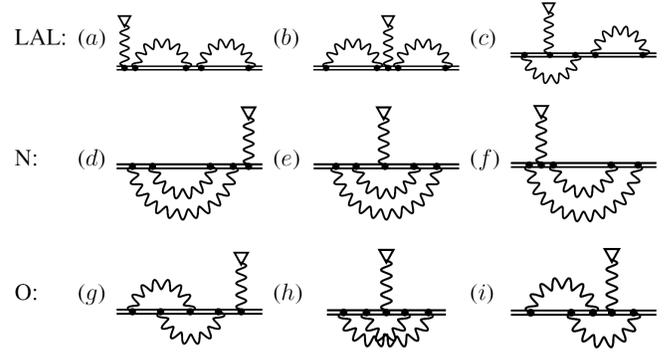}
\caption{Furry-picture diagrams of two-loop SE corrections to the $g$ factor. Double lines
represent electron wave functions or propagators, and the wave line represents a virtual photon.
A wave line terminated by a triangle represents the interaction with the magnetic field. Loop-after-loop (LAL), nested loop (N) and
overlapping loop (O) diagrams are shown in the respective rows.}
\label{fig:gSESE}
\end{figure}

There are three two-loop SE diagrams contributing to the \textit{binding energy} of a hydrogenlike ion, namely, the loop-after-loop~(LAL),
the nested loops~(N) and the overlapping loops~(O) diagrams. Their calculation has been presented in detail in
Refs.~\cite{Yerokhin2003,Yerokhin2003PRL,Yerokhin2006,Yerokhin2010,Mallampalli1998,Mallampalli1998PRL}.
The corresponding diagrams for the $g$ factor can be generated by magnetic vertex insertions
into the Lamb shift diagrams, yielding three nonequivalent diagrams in each of the above classes, shown in Fig.~\ref{fig:gSESE}.

\textit{Basic analysis.---}
We derived formulas for energy shifts induced by each diagram using the two-time Green's function formalism~\cite{Shabaev2002Report}.
The corresponding $g$-factor contribution $\Delta g$ is related to the energy shift by $\Delta E = - \frac{e B}{4 m_e} \Delta g$ (in relativistic units),
where $e$ and $m_e$ are the electron's charge and mass, respectively, and $B$ is the magnetic field strength.

We begin our analysis with the N and O diagrams. The diagrams with the magnetic field acting on one of the electron propagators inside the SE loops
(diagrams Fig.~\ref{fig:gSESE} $(e)$, $(f)$, $(h)$ and $(i)$) are called vertex diagrams. There are two types of electron
propagators the magnetic field can act on: following the nomenclature of Ref.~\cite{Yerokhin2003}, we call a vertex diagram \enquote{ladder} contribution
if the magnetic field acts on the central electron propagator [Fig.~\ref{fig:gSESE}~$(e)$ and $(h)$], and \enquote{side} contribution if the magnetic
interaction is connected with the leftmost or rightmost electron propagator [Fig.~\ref{fig:gSESE}~$(f)$ and $(i)$]. The energy shifts corresponding
to these diagrams can be written as
\begin{align}
\Delta E_{{\rm ver},i,j} = \langle a \vert \gamma^0 \Gamma^\mu_{ij} e A_{\mu} \vert a \rangle \,.
\end{align}
Here, $i \in \{ \mathrm{N,O} \}$ and $j \in \{ \mathrm{side,ladder} \}$, $\vert a \rangle$ denotes the $1s$ reference state,
$\gamma^0$ is the time-like Dirac matrix, $A_{\mu}$ is the magnetic four-potential with the Lorentz index $\mu$,
and the $\Gamma^\mu_{ij}$ are the two-loop vertex functions. The formulas for the latter are lengthy and will be
presented elsewhere.

The N and O diagrams in which the magnetic field acts on an external line [Fig.~\ref{fig:gSESE}~$(d)$ and $(g)$] need to be divided into
two parts. The electron propagator between the magnetic interaction and the SE loops can be represented as a sum over the spectrum of the Coulomb-Dirac
Hamiltonian, $G (E_a) = \sum_n \frac{|n\rangle \langle n | }{E_a-E_n(1-i0)}$, with the $E_n$ being eigenenergies of the eigenstates $|n\rangle$.
The cases $E_n \neq E_a$ and $E_n = E_a$ need to be analyzed separately. Following the usual convention in the literature (e.g. \cite{Beier00,Yerokhin04}),
we call these two contributions the irreducible (\enquote{irred}) and the reducible (\enquote{red}) parts, respectively. The energy shifts corresponding
to these diagrams are
\begin{align}
\Delta E_{i,\, \rm irred} = & 2 \langle a \vert \gamma^0 \Sigma_{i} \vert \delta_{\rm B} a \rangle, \\
\Delta E_{i,\, \rm red} = & \Delta E_{\rm mag} \langle a \vert \gamma^0 \left. \frac{\partial \Sigma_{i}}{\partial E} \right|_{E_a} \vert a \rangle\,.\nonumber
\end{align}
Here, the $\Sigma_i$ are the two-loop SE functions which are discussed in detail in Ref.~\cite{Yerokhin2003}.
$\vert \delta_{\rm B} a \rangle$ is the wave function perturbed  by the magnetic field, given as
$\vert \delta_{\rm B} a \rangle = \sum_{n \neq a} \frac{\vert n \rangle \langle n \vert - e \boldsymbol{\alpha} \cdot \boldsymbol{A} \vert a \rangle}
{E_a - E_n(1-i0)}$, with $\boldsymbol{\alpha}$ being the usual 3-vector of Dirac matrices. A closed expression for
$\vert \delta_{\rm B} a \rangle$ is known~\cite{ShabaevVirial}. $\Delta E_{\rm mag}$ is the energy shift corresponding to the
leading $g$-factor diagram~\cite{Breit1928}.

\begin{figure}
\includegraphics[width=\columnwidth]{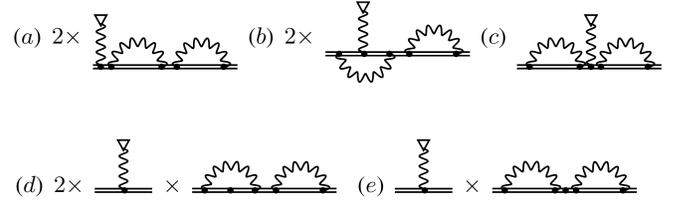}
\caption{Two-loop SE diagrams with the loop-after-loop structure (\enquote{LAL, irred} contribution). Double lines between two SE loops or between a
SE loop and the magnetic interaction represent here reduced Green's functions. A dot on an electron propagator denotes a derivative
with respect to the energy: $\left. \frac{\partial G (E)}{\partial E} \right|_{E=E_a}$.}
\label{fig:LALdiagrams}
\end{figure}

\begin{figure}
\includegraphics[width=\columnwidth]{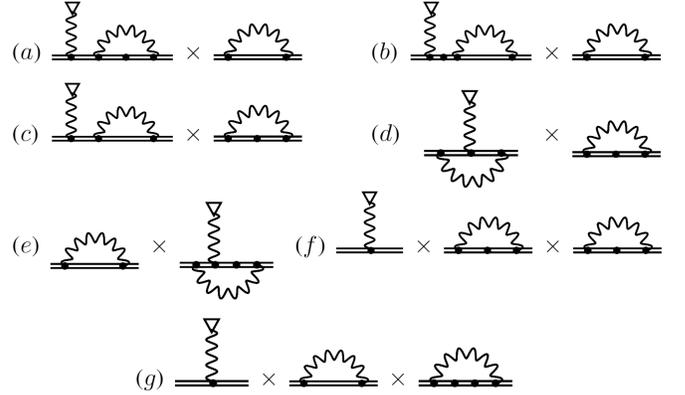}
\caption{Reducible two-loop SE diagrams which can be represented as products of one- or zero-loop diagrams (\enquote{LAL, red} contribution).}
\label{fig:LALreddiagrams}
\end{figure}

The LAL diagrams [Fig.~\ref{fig:gSESE} $(a)$ to $(c)$] give a large variety of contributions. In diagram \ref{fig:gSESE}~$(c)$, a separation
into the irreducible and the reducible part needs to be made for the propagator between the two SE loops, similarly to the case of the N and O
diagrams. The reducible part can be represented as a product of two one-loop functions, and the irreducible part consists of two one-loop functions
connected by a reduced Green's function $G_{\rm red} (E_a) = \sum_{n, n \neq a} \frac{|n\rangle \langle n | }{E_a-E_n(1-i0)}$. In diagrams
\ref{fig:gSESE}~$(a)$ and $(b)$, there are two propagators for which this separation needs to be made.
We therefore distinguish between the cases of $E_n \neq E_a$ for both propagators (\enquote{irred, irred}), $E_n \neq E_a$ for one propagator and
$E_n = E_a$ for the other propagator (\enquote{irred, red}) and $E_n = E_a$ for both propagators (\enquote{red, red}). The \enquote{irred, irred}
contributions consist of diagrams with two SE loops connected by $G_{\rm red}$. The \enquote{red, red} contributions can be represented
as products of three diagrams, namely the leading order $g$-factor diagram and two one-loop diagrams. Finally, there are two kinds of
\enquote{irred, red} contributions. First, there are contributions which can be represented as products of two one-loop diagrams. Second, there are
contributions which can be represented as a product of the leading-order $g$-factor diagram and a diagram which contains two SE loops connected by
$G_{\rm red}$. We cast all LAL contributions into the \enquote{LAL, irred} and the \enquote{LAL, red} categories, shown in Figs.~\ref{fig:LALdiagrams}
and \ref{fig:LALreddiagrams}, respectively.

\begin{figure*}[t]
\includegraphics[width=1.3\columnwidth]{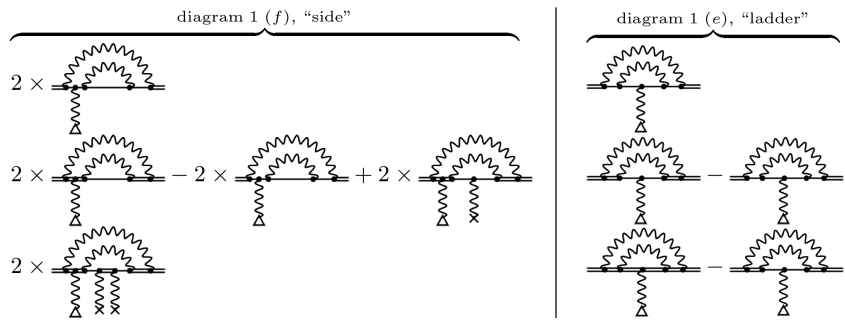}
\caption{Separation of the \enquote{N, vertex} diagrams [Fig.~\ref{fig:gSESE}~$(e)$ and $(f)$] into F- (first line), P- (second line) and M-term
(third line). A wave line terminated by a cross represents a Coulomb interaction.}
\label{fig:NverFPM}
\end{figure*}

\textit{Regularization of divergences.---}
Self-energy corrections suffer from ultraviolet (UV) divergences, which have to be separated carefully~\cite{Yerokhin2003}. The standard
renormalization method has been elaborated in momentum space for diagrams containing free Dirac propagators, while the Coulomb-Dirac propagators
are only known in coordinate space. Therefore, in our regularization scheme we subtract diagrams with the Coulomb-Dirac propagators replaced
by propagators containing zero or one interaction with the Coulomb field in such a way that the corresponding difference is rendered UV finite.
The subtracted diagrams can then be evaluated in momentum space or in a mixed momentum-coordinate representation. In case of the one-loop Lamb
shift or the one-loop SE correction to the $g$ factor, this approach was implemented in Refs.~\cite{Snyderman91,Yerokhin99,Yerokhin04}.
In case of the two-loop SE correction to the $g$ factor, one encounters overlapping UV divergences. E.g., the O SE function in diagram
\ref{fig:gSESE}~$(g)$ consists of two overlapping one-loop vertex functions, each of which give rise to UV divergences. This property renders
the isolation of divergences much more cumbersome.

Furthermore, infrared (IR) divergences may appear whenever the energy of an intermediate state coincides with $E_a$ \cite{Yerokhin2003}. Such
reference-state IR divergences are present in the one-loop $g$-factor correction as well as in the two-loop Lamb shift.
In both cases, it is possible to identify diagrams which are each IR divergent on their own but whose sum is IR finite. The situation for the two-loop
SE correction to the $g$ factor is more complicated, requiring an adequate regrouping of different terms. Our analysis of divergences
shows a partial cancellation of UV and IR divergences between the different N and O diagrams. The remaining UV and IR divergences in the N and O diagrams
are cancelled exactly by the divergences in the \enquote{LAL, red} contribution. The \enquote{LAL, irred} contribution is both UV and IR finite.

\textit{Separation into categories.---}
In order to handle divergences, we split all diagrams into different categories. One-loop functions can be split into the zero-, one-
(if necessary), and many-potential terms. The zero- and, in some cases, the one-potential contributions are UV divergent. These divergent
contributions are evaluated in momentum space, using the dimensional regularization procedure~\cite{QFT}. The many-potential functions which are UV
finite are computed in coordinate space, as these involve the Coulomb-Dirac propagator. The \enquote{LAL, irred} and the \enquote{LAL,~red}
contributions are dealt with using a straightforward generalization of this procedure.

The situation is more complex for the N and O diagrams. While in the one-loop case, diagrams can always be divided into
UV-divergent terms, and contributions which contain the Coulomb-Dirac propagator, two-loop diagrams need to be divided into three different
categories: (i) diagrams which contain UV divergences, (ii) diagrams which contain the Coulomb-Dirac propagator and (iii) diagrams which contain both.
Using the nomenclature introduced for the two-loop Lamb shift, we refer to these categories as the F-, M-, and P-term,
respectively \cite{Mallampalli1998}.

\begin{figure}
\includegraphics[width=1.0\columnwidth]{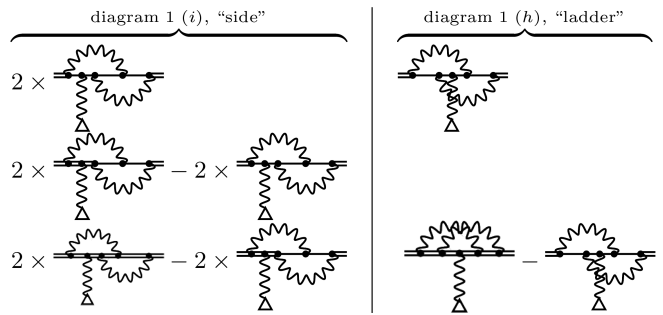}
\caption{Separation of the \enquote{O, vertex} diagrams [Fig.~\ref{fig:gSESE}~$(h)$ and $(i)$] into F- (first line), P- (second) and M-term (third line).
Note that the \enquote{ladder} diagram \ref{fig:gSESE}~$(h)$ does not contain a P-term contribution because there is no divergent subgraph in this diagram.}
\label{fig:OverFPM}
\end{figure}

Replacing $\vert \delta_{\rm B} a \rangle$ with $\vert a \rangle$ in the \enquote{N, irred} and \enquote{O, irred} diagrams [Fig.~\ref{fig:gSESE} $(d)$
and $(g)$], one obtains the known Lamb shift contributions. Therefore, the separation of these diagrams into F-, M- and P-terms is identical to the case
of the Lamb shift~\cite{Yerokhin2003}. For the N and O reducible and vertex diagrams, we consider the expansion of the electron propagators in powers of
the interactions with the nuclear potential and analyze the superficial degree of divergence $d$, as defined in \cite{QFT}.
We divide the contributions into F-, P- and M-terms according to the definitions\\[1mm]
\begin{tabular}{lll}
\quad\quad & $d \geq 0$:                        & F term, \\
\quad\quad & $d < 0$, UV-divergent subgraph:    & P term, \\
\quad\quad & $d < 0$, no UV-divergent subgraph: & M term.
\end{tabular}\\[1mm]
The separation of the \enquote{N, vertex} and \enquote{O, vertex} diagrams is illustrated in Figs.~\ref{fig:NverFPM} and \ref{fig:OverFPM},
respectively. The \enquote{O, red} and the \enquote{N, red} diagrams can be treated analogously.

\textit{Numerical results.---}
In order to assess the relevance of a non-perturbative theory, we evaluate first the F term. This term is expected to be the dominant one as it
incorporates the free-electron two-loop SE correction. The calculation typically involves the evaluation of matrix elements of two-loop SE functions
which are partially known~\cite{Yerokhin2003}, or in the case of a magnetic insertion, were derived in the current work. Matrix elements are calculated
either with Coulomb-Dirac wave functions in coordinate or momentum space, or with the wave function $\vert \delta_{\rm B} a \rangle$.
Complex $\gamma$ matrix expressions were reduced by computer algebraic methods~\cite{Mathematica}. Feynman integralss were
either carried out analytically, again with the help of symbolic computing~\cite{Mathematica}, or numerically, employing standard or the recently
developed extended Gauss-Legendre quadratures~\cite{gausext}. We tested our numerical codes by replacing $\vert \delta_{\rm B} a \rangle$
with the regular bound-electron wave function in certain diagrams, reproducing known Lamb shift contributions \cite{Yerokhin2003}.

For the free-electron case, i.e. in the limit of an infinitesimally weak Coulomb potential, all P and M terms and the one-potential
F terms vanish. Furthermore, we expect all \enquote{LAL, irred} contributions to converge to zero, as well as those
\enquote{LAL, red} diagrams which contain the one-loop SE correction [Fig.~\ref{fig:LALreddiagrams}~$(a)$, $(b)$, $(e)$, and $(g)$], or the irreducible
one-loop SE wave function correction to the one-loop $g$ factor [Fig.~\ref{fig:LALreddiagrams}~$(c)$] as a factor.

We define the total zero-potential F-term contribution to consist of the N and O vertex and reducible diagrams, the zero-potential contributions
(of both factors) to the \enquote{LAL, red} diagrams \ref{fig:LALreddiagrams}~$(d)$ and $(f)$, and the irreducible zero-potential N and O contributions.
The reducible F-term contribution consists of the remaining \enquote{LAL, red} diagrams with free internal lines. The one-potential F-term
consists of the irreducible one-potential N and O contributions. Numerical values and their uncertainties are given in
Table~\ref{tab:FLALZalpha}.

\begin{figure}[t!]
\includegraphics[width=\columnwidth]{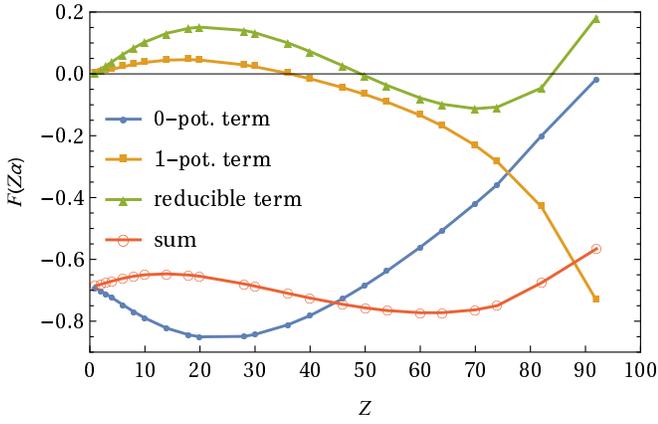}
\caption{The zero-, one-potential and the reducible F term contributions as a function of $Z$,
expressed in terms of the scaled function $F(Z \alpha)$, defined as $\Delta g = \left(\frac{\alpha}{\pi}\right)^2 F (Z \alpha)$.
See text for further details.}
\label{fig:F-all-vs-Z}
\end{figure}

According to the above discussion, we expect the sum of the zero-potential F-term contribution to converge to the free-electron two-loop SE
correction for $Z~\to~0$. The free-electron $g$-factor contribution can be determined using the form factors \cite{Pachucki05}, and our results
converge well to this value in the low-$Z$ limit (see Fig.~\ref{fig:F-all-vs-Z} and
Table~\ref{tab:FLALZalpha}). Fig.~\ref{fig:F-all-vs-Z} shows a complex dependence of the calculated F terms on the atomic number $Z$, which largely
deviates from the result of the $\Za$ expansion up to fourth order, highlighting the need for a non-perturbative-in-$\Za$ theory.

\begin{figure}[t!]
\includegraphics[width=\columnwidth]{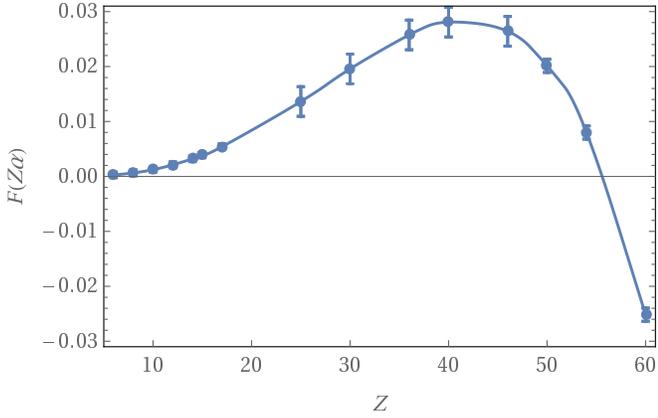}
\caption{The complete \enquote{LAL, irred} contribution, as a function of $Z$, given in terms of the scaled function $F(Z\alpha)$.
}
\label{fig:FLAL}
\end{figure}

In the case of the Lamb shift, the LAL correction gives an estimate of the total two-loop SE correction in a wide $Z$ range
\cite{Yerokhin2003,Mallampalli1998PRL}. To check whether this also holds in our case, we evaluate the LAL $g$-factor contributions.
To this end, it is convenient to define a \enquote{SE-perturbed wave function} (see e.g. \cite{Holmberg15})
$\vert \delta_\Sigma a \rangle = \sum_{n , n \neq a} \frac{\vert n \rangle \langle n \vert {\Sigma}_{\rm R} \vert a \rangle}{E_a - E_n}$,
with the regularized one-loop SE operator ${\Sigma}_{\rm R}$. The most difficult aspect of the LAL calculation is the numerical determination of
$\vert \delta_\Sigma a \rangle$, for which we used the B-spline method \cite{JohnsonBSplines,Shabaev2004}.
The $g$-factor contribution corresponding to diagram \ref{fig:LALdiagrams}~$(e)$ is:
$\Delta g_{\rm LAL,e} = - g_{\rm D} \langle \delta_\Sigma a \vert \delta_\Sigma a \rangle$. The computation of diagram \ref{fig:LALdiagrams}~$(c)$
is similar to the formula for the computation of the Dirac value $ g_{\rm D} = - \frac{8 m_e}{3} \int \limits_0^\infty \mathrm{d} r r^3 f_a(r) g_a(r) $,
with the radial components $g_a(r)$, $f_a(r)$ of the usual wave function replaced by those of $\vert \delta_\Sigma a \rangle$.
The remaining LAL diagrams can be rewritten as matrix elements of the one-loop SE or vertex functions, with either the usual wave function or
$\vert \delta_{\rm B} a \rangle$ on one side and $\vert \delta_\Sigma a \rangle$ on the other side. The one-loop operator has to be expanded into
\mbox{zero-,} one- (if necessary), and many-potential terms. Numerical results for the total \enquote{LAL, irred} contribution are given in the last
column of Table~\ref{tab:FLALZalpha} and shown in Fig.~\ref{fig:FLAL}. Unlike in case of the Lamb shift, in the $g$~factor the F terms dominate due to
the nonvanishing free-electron limit up to high $Z$ values. Fig.~\ref{fig:FLAL} demonstrates that the behavior of the \enquote{LAL, irred} term at
intermediate and high $Z$ largely deviates from its low-$Z$ characteristics, and it even changes sign around $Z=55$. A highly nonperturbative behavior
of the LAL term was also observed in Lamb shift calculations~\cite{Mallampalli1998PRL}.

\textit{Summary.---}
The theoretical framework for the evaluation of two-loop SE corrections to the $g$ factor in a non-perturbative nuclear
field has been developed. The isolation of divergences was carried out by separating the LAL, N and O Furry-picture diagrams
into terms consisting of diagrams with UV divergences, diagrams which contain a Coulomb-Dirac propagator, and diagrams which contain both.
Such a rearrangement assures finite results. Numerical results are given for the dominating group of terms, the F terms, namely, those in which
interaction of the nucleus in the intermediate states is treated up to first order, and for the LAL diagrams. The results show that
a non-perturbative treatment is essential in a rigorous description of the bound-electron $g$ factor, and will be relevant to projected
tests of QED in strong Coulomb fields and to the determination of $\alpha$~\cite{Sturm17,Shabaev06}
in planned experiments with highly charged ions~\cite{Sturm17}.

\begin{table}[t]
\begin{center}
\begin{tabular}{cllll}
\hline\hline
 $Z$ & $F_{\rm F,0pot}$ & $F_{\rm F,red}$         & $F_{\rm F,1pot}$         & $F_{\rm LAL}$           \\
\hline
 1   & -0.693181(19)    & \phantom{-}0.005715(2)  &  \phantom{-}0.00213(27)  &       --                \\
 2   & -0.701989(10)    & \phantom{-}0.015596(2)  &  \phantom{-}0.00576(27)  &       --                \\
 3   & -0.712496(9)     & \phantom{-}0.026977(2)  &  \phantom{-}0.01011(18)  &       --                \\
 4   & -0.723816(6)     & \phantom{-}0.038885(2)  &  \phantom{-}0.014544(44) &       --                \\
 6   & -0.747062(4)     & \phantom{-}0.062437(2)  &  \phantom{-}0.023242(27) & \phantom{-}0.00026(53)  \\
 8   & -0.769444(6)     & \phantom{-}0.084105(2)  &  \phantom{-}0.030852(6)  & \phantom{-}0.00064(53)  \\
 10  & -0.789865(4)     & \phantom{-}0.103024(2)  &  \phantom{-}0.037022(16) & \phantom{-}0.00123(53)  \\
 30  & -0.842748(4)     & \phantom{-}0.132908(9)  &  \phantom{-}0.023105(1)  & \phantom{-}0.0196(27)   \\
 40  & -0.781697(4)     & \phantom{-}0.072266(9)  &            -0.015710(1)  & \phantom{-}0.0281(27)   \\
 50  & -0.683620(4)     &           -0.006450(14) &            -0.066763(1)  & \phantom{-}0.0201(12)   \\
 60  & -0.560628(4)     &           -0.077985(13) &            -0.134018(2)  & -0.0252(12)             \\
 70  & -0.419265(4)     &           -0.112472(14) &            -0.231468(2)  & -0.1405(12)             \\
 92  & -0.016259(15)    & \phantom{-}0.183186(32) &            -0.732700(4)  & -0.9734(39)             \\
\hline\hline
\end{tabular}
\caption{The zero-, one-potential and reducible F and LAL term contributions for different atomic numbers.
For $Z \to 0$, $F_{\rm F,0pot}$ F-term converges to the free-electron limit $F(0) = -0.68833\dots$.}
\label{tab:FLALZalpha}
\end{center}
\end{table}

\begin{acknowledgments}
This work is part of and supported by the German Research Foundation (DFG) Collaborative Research Centre "SFB 1225 (ISOQUANT)".
V.~A.~Y. acknowledges  support by Ministry of Education and Science of Russian Federation (grant No. 3.5397.2017/6.7).
\end{acknowledgments}

\end{document}